# Voice over IP in the Local Exchange: A Case Study[*]

**Dr. Martin B. H. Weiss**
Telecommunications Program
University of Pittsburgh
email: mbw@tele.pitt.edu
http://www2.sis.pitt.edu/~mweiss

**Hak Ju Kim**
Telecommunications Program
University of Pittsburgh
email: hjkim@mail.sis.pitt.edu
http://www2.sis.pitt.edu/~hjkim

**27 August 2001**

## ABSTRACT

There have been a small number of cost studies of Voice over IP (VoIP) in the academic literature.  Generally, they have been for abstract networks, have not been focused on the public switched telephone network, or they have not included the operating costs. This paper presents the operating cost portion of our ongoing research project comparing circuit-switched and IP network costs for an existing local exchange carrier.

We have found that (1) The operating cost differential between IP and circuit switching for this LEC will be small; and (2) A substantial majority of a telco's operating cost lies in customer service and outside plant maintenance, which will be incurred equally in both networks in a pure substitution scenario.  Thus, the operating cost difference lies in the actual cost differences of the switching technologies.  This appears to be less than 10%-15% of the total operating cost of the network. Thus, even if the cost differences for substitute services were large, the overall impact on the telco's financial performance would be small. But IP has some hidden benefits on the operations side. Most notably, data and voice services could be managed with the same systems infrastructure, meaning that the incremental operations cost of rolling out new services would likely be much lower, since it would all be IP.

*Key Words: Circuit-switched and IP (Packet-switched) network, Top-down and Bottom-up Cost Approach, LRAIC, USOA Accounting Rules, DMS SuperNode, Softswitch, VoIP, HAI model.*

## 1. INTRODUCTION

Voice over IP has garnered substantial attention from the trade press, regulators, and industry analysts.  Some have asserted that VoIP is a form of regulatory arbitrage, implying that it will disappear as soon as regulatory treatment of different communications forms disappears.  Others

---

[*] This work was funded by a grant from the Internet and Telecommunications Convergence Consortium

claim that VoIP demonstrates the characteristics of a "disruptive technology" [30], suggesting that incumbent firms are at risk. If it is the latter, then its cost structure, its features and capabilities, and its performance must prove to be superior in combination to the current generation technology (circuit switching).

Still other industry observers have argued that declining prices of mobile wireless services will reduce revenues from voice services, forcing LECs to seek data traffic to support bottom line growth. If this analysis is correct, it suggests that it may be efficient for LECs to convert their switching technology to packet from circuit.

This possible future scenario has raised the interest of academic and industry researchers. Their attention has been focused not only in the technological problems with such a transition, but also on understanding its cost implications. Most of the cost studies have been focused on the initial investment (capital cost) by comparing to the existing phone network (PSTN). Further, most of cost studies of VoIP have been for abstract networks, like "green field" networks and have not included the operating costs. While these studies have been useful in understanding the first order cost factors, the "green field" assumption has limited usefulness in reality; further, understanding and controlling operating costs is becoming as critical as equipment costs because of the ever-increasing level of competition. If a network is already in place, the initial investment is typically considered a sunk cost, and therefore irrelevant for the comparison of Circuit-switched and IP network.

A previous study by Weiss, Kim, and Hwang[1] found that a VoIP network could reduce the required link capacity (hence reducing cost); further, the switching equipment costs of a circuit switched network were estimated to be higher than an IP-based network. In this paper, we consider operating costs in more detail, not just with an abstract network, but using a 75,000 subscriber LEC network as a basis, and a scaled-up 30,000 node IP network of the University of Pittsburgh. The underlying network data are derived from our studies of the LEC's and the University of Pittsburgh's network. Actually direct comparison of two networks may be not entirely fair, since current circuit-switched network tends to incorporate many capabilities and features that are not implemented in IP network yet. However, the basic points (e.g. switching functions) may be compared.

Comparing these two network technologies is challenging, as they provide substantially different features and capabilities. For this research, we assume that the VoIP network provides exactly the same services as the circuit switched network, no more, no less. With this, we assume that the major operating cost components, namely sales, billing and customer support, are invariant, as are all non-switching costs (eg., outside plant). The focus of our investigation, then is on the operating costs associated with the switching and signaling functions.

This study followed the module-based approach shown in Figure 1-1. The first module studies the network evolution trends including softswitch & broadband services, and also identifies motivation &

scope of our study. In the second module, we reviewed the existing cost models of PSTN and IP network. The third module deals with the network related issues such as network architecture, components, and operation & management (OAM)) based on actual existing networks: an ILEC and PittNet (University of Pittsburgh IP network). The operating costs of circuit and IP network will be analyzed in the fourth module.

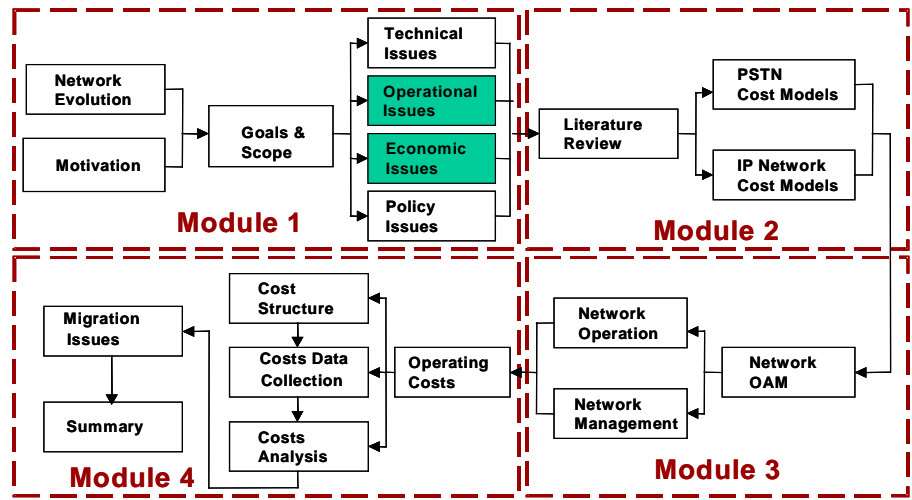

**Figure 1-1 Flow-Chart of the Study**

## 1.1 Network Evolution

To further motivate this paper, it is useful to briefly describe the context in which this potential transition will take place.  Much of this is driven by a vision of the future in which multimedia applications require voice, video, and data in combinations that are just being discovered by the industry.  In telecommunications, technical convergence refers to a single technical infrastructure that can handle all of these streams on a single network rather than on multiple special purpose networks.[2-4]

### 1.1.1 Broadband Service

Another key driver for the future network is the provision of broadband services.  The most costly aspect of this by far is the so-called last mile.  .  Thus, the local access network is often the bottleneck for the rollout of broadband services.[5,6] By 1999, consumers and businesses had an insatiable demand for high-speed access to the Internet and various other networks.[5] The solutions that emerged for broadband access are DSL (digital subscriber line) [7], cable modem and satellite [4]. These technologies can support high-speed data services [8] (e.g. broadband services) and access directly into the Internet without having to use the conventional telephone network (PSTN).

Since the existing central office infrastructure (voice-centric network) represents a large investment, the local exchange carriers (LECs) are unlikely to replace it with a data-centric network any time soon. Instead, the PSTN has been evolving from basic circuit-switching functionality to include more advanced network capabilities based on the advanced intelligent network (AIN). [9,10] The PSTN is gradually becoming more data-centric with high-speed access (DSL in the local loop) and deployment (optical transport by wave division multiplexing).

**1.1.2 Switching Architecture**

As IP-based technologies have evolved, network users have found new uses for them: for example, IP technologies can transmit video as well as voice. The trend to carry voice telephony connections over the Internet has become one of the major revolutions in the modern telecom industry. Many companies including traditional telcos have made large investments towards implementing high-speed packet technologies in their networks, especially Class 3 and 4 switching.[10]

This is evident in the evolution of the "softswitch" technology.  Here, network hardware is separated from network software [9,11,12] (see Figure 1-2). In traditional circuit switching equipment, hardware and software are not independent; they rely on dedicated facilities for interconnection and are designed primarily for voice communications. Packet based networks, such as those that use the Internet Protocol (IP) , route voice and data over diverse routes and shared facilities. [9] As seen Figure 1-2, Media gateways are an elaboration of the "gatekeeper", which is derived from early VoIP systems in which gateways converted the voice and signaling from analog PSTN and SS7 to IP packets. The gateways use a protocol called media gateway control protocol (MGCP), and they do not have to be collocated with the softswitch. The softswitch sets up the path or does a database dip, but the gateway controls the flow of the medium -- whether it is voice or data -- over the path.[9]

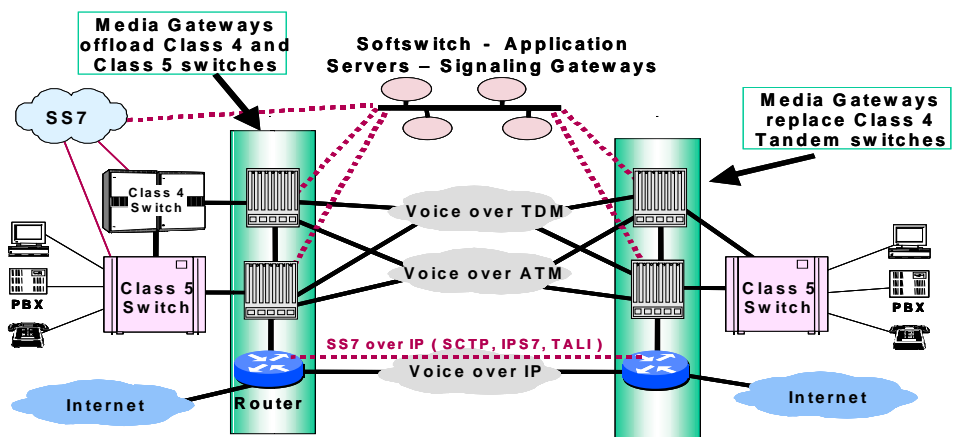

**Figure 1-2 A Converged Network of PSTN and IP network,** *Source: RHK, 2001*

## 2. COSTING MODELS: AN OVERVIEW

### 2.1 PSTN Cost Models

Several approaches have been used to estimate network costs in the traditional telephone network (PSTN), but the two most common approaches are the top-down approach and the bottom-up approach.[13].[14,15]  A top-down approach (or accounting approach) uses the actual accounting data of an operator, and allocates costs to different services based on the relationship between costs and services. Typical of this approach is the FCC Part 36 rules (jurisdictional separations procedures) based on the Unified System of Accounting (USOA).

In contrast, a bottom-up approach (or engineering/economics approach) involves the development of engineering-economic models in order to calculate the costs of the network elements required to provide particular services, assuming modern technology and efficient methods of operation. The typical example is a COnstructive COst MOdel (COCOMO), which is a well-known model used in software cost and schedule estimation.[16]

These two approaches can be used together in a hybrid (or mixed) approach.  An example of this is the long run average incremental costs (LRAIC) method.[16] In this method, the accounting data provided by the operator is used to value current equipment according to a current cost method; this includes the predictable network upgrades according to a pre-determined plan.  Meanwhile, a technical-economic model is used to design an optimal network that an operator *could* build using the best available technology to meet the current demand.

An example of LRAIC is found in the FCC's First Order [17] implementing the Telecommunication Act of 1996, when they specified the TELRIC method. This is a specially designed cost methodology that draws on accounting data (a top-down approach), economics & engineering information (a bottom-up approach) to perform cost analysis. The engineering economics provides a fully costed, forward-looking incremental cost of service, and the accounting data provides the asset and expense information classification system. In a sense, TELRIC is a traditional fully distributed cost methodology, but it distributes forward-looking costs, not actual cost from accounting records (see, for example the HAI Model.[18]).

## 2.2 IP/VoIP Network Cost Models

Since Internet Service Providers (ISPs), unlike traditional telcos, are not regulated, no uniform formal cost model exists (such as the FCC's). At present, several cost models concerning ISPs' network and VoIP network are introduced as follows.

### 2.2.1 Mackie-Mason & Varian's Model

Mackie-Mason & Varian [19] identify five types of costs that occur in computer networks (IP network). First is a fixed cost of providing the networking infrastructure. Second is an incremental cost of connecting to the network, which is usually paid by the user in the form of a connection cost. Third is the cost of expanding the network's capacity (congestion cost). Fourth is the incremental cost of sending an extra packet. Finally, the transmission of a packet invariably leads to a delay in other users' packets. This externality should be considered a social cost. A number of these costs are analyzed for the Internet in [20].

### 2.3.2 Leida's VoIP Model: ITSP Model

Leida [21] used a hybrid approach, which is mixed top-down approach and bottom-up approach. That is, the model used an ISP's present cost (top-down approach) as an IP network costs and quantified the impact of Internet telephony on these costs (bottom-up approach).

The model used the five types of subscribers, which are residential dial-in, business dial-in, 128 kb dial-in ISDN, 56 kb leased-line, and T1 leased-line subscribers. Hence, the ISP's overall cost distribution varied substantially according to the mix of subscribers. The model considers five cost categories of an ISP: capital equipment, transport, customer service, operations and other expenses (sales, marketing, general and administrative). Each category in detail is shown in Table 2-1.

| Categories | Cost Items |
|---|---|
| Capital Equipment | Analog modems, content housing server, 56 kb CSU/DSU, Cisco 7513 serial port card, 4 port DS3 card, etc. |
| Transport | Leased lines to connect the Tier 1 and Tier 2 POPs (T1, T3 and OC-3s), incoming analog and ISDN phone lines (T1 & PRI), monthly costs for the ISP to NAP, etc. |
| Customer Service | Staff and facilities for supporting the customers, e.g., technical support to the subscribers, etc. |
| Operations | Billing, facilities maintenance, operations personnel, etc. |
| Other Expenses | Sales/marketing, general/administrative expenses, etc. |

**Table 2-1 Cost Categories**

### 2.3.3 CMIT (Cost Model for IP Telephony) Model

The CMIT model [22] was submitted to FCC by the students of the telecommunications modeling and policy seminar in 1996. It assumes three IP Telephony scenarios: IP Telephony between computers (Internet Access Provider model), IP Telephony between telephones (gateway model), and IP Telephony assumed local access charge (regulatory model). The costs categories are local loop costs, Internet point-of-presence (IPOP) costs, and upstream costs.

The local loop consists of local distribution lines and an interoffice line; it accounts for multiplexing at the central office between the residence and the Internet Access Provider (IAP). However, the local loop does not account for the lines or hookup costs between residences and the central office. The gateway costs are the sum of the internal cost, obtained from the analysis of a currently existing IPOP, and possibly the additional hardware cost for gateway implementation. Internal cost consists of the costs for incoming and outgoing trunk lines depending on the cumulative bandwidth requirements (T1/T3) and the distance from the central office. The upstream costs consist of the interconnection between the ISP and the national access provider (NAP), as well as the Interconnection between the various NAPs.

### 2.3.5 ICS (Internet Cost Structure) Model

Srinagesh[23] categorized ISPs' costs as follows: hardware & software, customer support, IP transport, and access to information sources on the Internet. As transport costs fall and firms seek to differentiate their services, support costs tend to rise as a fraction of total cost. There are many costs associated with offering all of the services listed in ISP, such as layer 2 transport (private lines, SMDS, Frame Relay, and ATM--approximately 30-40% of total costs), network management (address assignment, routing, etc.), customer support (marketing, etc.), information acquisition, and general management.

### 2.3.6 ISPC (ISP's Cost Model using Gateway) Model

Andrew Sears [24] assumed that an ISP has a telephone gateway. So, if ISPs offer voice access across the Internet, it is likely that the costs to do so would be similar to the cost of providing data access. ISPs could simplify billing and combine advertising costs by offering their customers a flat rate bundle for unlimited voice and data access. Using this type of model, many of the variables in the cost function of providing data and voice access to the Internet would be the same.

The cost variables include the server, telephony/modem cards, billing, labor, space rental and advertising. The most significant cost would be the cost of local lines. Since each phone call involves two servers, the cost of each voice call could be as the cost of two data access lines for an ISP plus the local dial out cost as shown in Figure 2.6.

$$C_{voice} = 2C_{data} + C_{local\ call}$$

\* $C_{voice}$ : Cost of Each Voice
\* $C_{data}$ : The Cost of Data Access Lines for ISP
\* $C_{local\ call}$ : The Local Dial Out Cost

**Figure 2.6 The ISPC Model**

Rather than performing a cost analysis of ISPs and phone gateways, the above cost equation was used to show that ISPs could profitably offer competitive prices compared to the prices of Interexchange Carriers (IXCs).

# 3. CASE STUDIES

## 3.1 General

In this section, we present the results of our studies of the two actual networks that form the basis of the conclusions of this research. As mentioned above, we collected detailed cost and configuration data from a 75,000 line ILEC and from the University of Pittsburgh's IP network. To compare these networks, we focus on per line (or per node) analysis. Note that this assumes that operating costs scale linearly; we have not tested this assumption.

In supporting its operations, the University of Pittsburgh maintains a circuit switched voice network and an IP-based data network. Even though PittNet separates voice and data handling, the IP portion can be useful to model an IP telephony network model because the IP network structure itself is does not change much when voice functionality is added. Later in costs analysis, these added costs are included as spare costs. So, we assume that PittNet can support telephony services over its IP packet infrastructure with the same call control and quality of service (QoS) across the network.

The ILEC that we study is capable of providing Digital Subscriber Line (DSL) throughout its service area. It also has a division that provides competitive local exchange service and another that sells equipment. Since we focus on the direct operational cost of switching, we ignore the problems of attributing the indirect costs of managing this multitude of divisions.

## 3.2 The ILEC Network

### 3.2.1 Network Description

The LEC we are using as our case is a medium sized independent telephone company. It provides local network services as an ILEC and has CLEC, telecom equipment sales, internet access and web hosting, and long distance. It has invested heavily in developing an infrastructure that is modern and capable of providing a wide array of telecommunications services. Like many independents at the beginning of the current competitive era, the LEC had a combination of old Step-by-Step and early generation digital switching systems; their switching plant was fully digital by the late 1980's.

Figure 3-1 shows the network architecture of the LEC. The main office was set up as a master unit (the DMS-100/200 switch: 100 for local, 200 as a toll tandem), and most of the outlying exchanges were Remote Switching Offices (RSCs and RLCMs). The LEC supported about 75,000 subscribers in 1999.

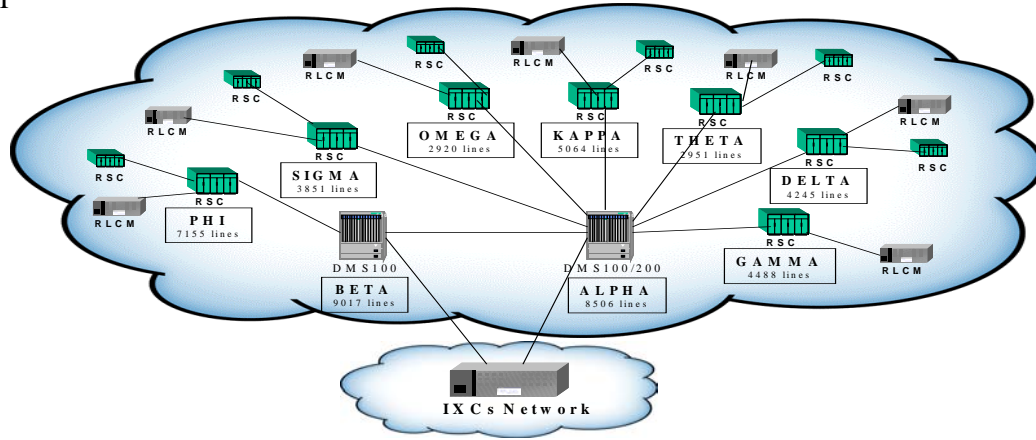

**Figure 3-1: LEC's Physical Network Architecture**

### 3.2.2 Central Office (CO)

Approximately 60 to 85 percent of the interior space in a telephone equipment building consists of equipment areas. These are large, usually windowless, partitionless rooms designed to contain the equipment, the appropriate cable support systems. The major functions of the CO are:

- To connect between users by central office switches (messaging)
- To route calls onto digital trunks (switching)
- To provide billing information by using call detail records (CDRs).
- To provide management information, such as alarming, usage, etc.

### 3.2.3 Switching System

### 3.2.3.1 Hardware

Figure 3-2 shows a block diagram of the architecture of the DMS SuperNode System that the LEC uses. The DMS SuperNode system is a digital time-division electronic switching system designed for layered and modular growth structure to accommodate local offices ranging from 10,000 to 135,000 lines.

It consists of a number of interface modules: Computing Module (CM), ENET switch matrix, Link Peripheral Processor (LPP), Peripheral Modules (PM), Remotes, and Operation Administration & Maintenance (OAM). The CM provides computing and memory resources for overall system management and control as the central processing engine of the DMS SuperNode system. The ENET is the switch fabric. The LPP supports a mix of up to 36 interface units, providing such services as: Signaling Network messaging, ISDN packet handling, Ethernet internetworking to the DMS Billing Server, Frame Relay support, Automated Directory Assistance Service (ADAS), etc.  The PMs are the bridge between the DMS system and the voice and data traffic over trunks and lines. These modular frames offer flexible connectivity to digital and analog transmission facilities, maintenance circuits, subsidiary PMs, or other switching offices. DMS Remotes extend the reach of DMS services significantly deeper into the network.  The OAM subsystems offer the management of the DMS node.

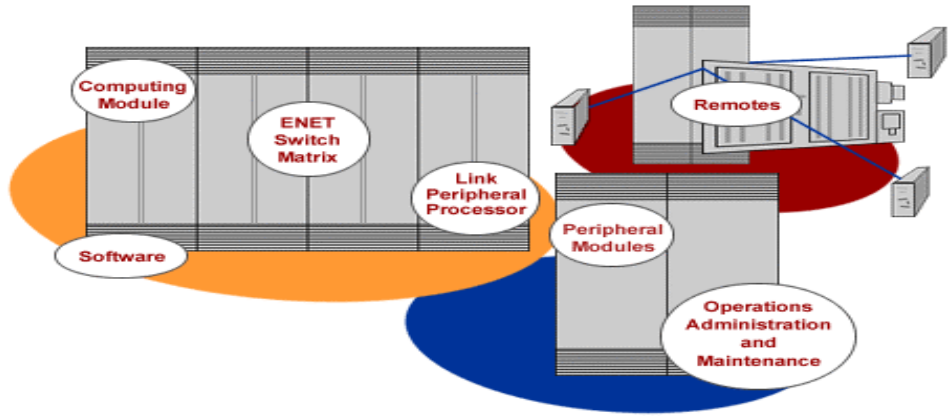

**Figure 3-2 DMS SuperNode System Architecture**
*Source: Nortel Network*

### 3.2.3.2 Software

The complexity of the switch software, coupled with the inherent lack of flexibility and control over its capabilities by the service providers, led to the inception and deployment of intelligent network (IN).[25] Traditionally voice-switching applications have been proprietary software running on switch hardware. That has typically meant long waits for service providers that wanted new applications to differentiate their services. They were basically at the mercy of their switch vendors, who could roll out new software enhancements at their leisure.

The DMS system software policy (PCL; product computing module load) is life-cycle system, which is upgrade periodically with two years circular periods. [26] Figure 3-3 shows the software life-cycle of the DMS SuperNode system. A DMS system is upgraded smoothly from an existing release to a more current production or active software release. For example, seen in Figure 3-3, a system can upgraded from one release to the next, or skip one release (release 1 to release 3), or skip two releases (release 1 to release 4).

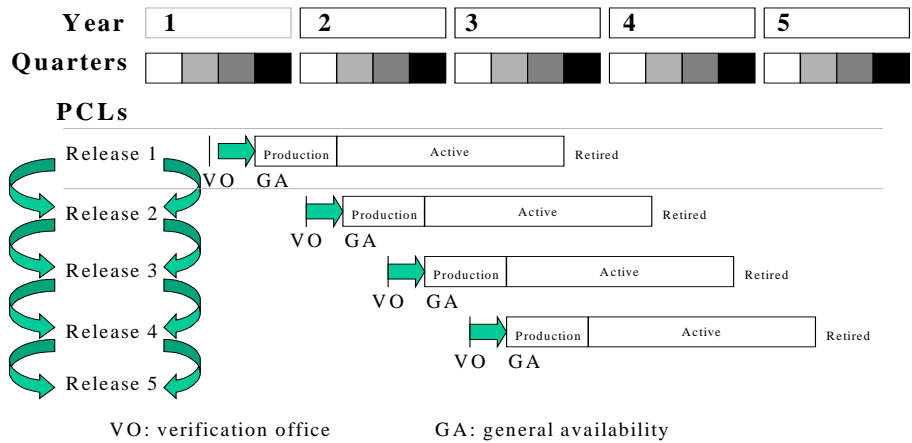

**Figure 3-3 Software Life-Cycle (*Source: Nortel Networks*)**

### 3.2.4 Network Operation & Management (OAM)

### 3.2.4.1 Network Operation

The evolution of network operation occurred mainly in the switching and transmission systems. For the most part, network operations systems were automated from operations that were previously performed manually by the telephone company employees. The generic network operation categories are: customer Services, service provisioning, network operations, planning & engineering, procurement.[27]

**(1) Customer Service Activity**

Customer service activity includes to receive customer requests and to assign them to related departments in the business office. Before assigning them, the department checks cable records and then write service order.

**(2) CPE Management Activity**

CPE management activity includes visits to the customer premises for the installation and repair department.

**(3) Service Provisioning Activity**

Service provisioning activity includes all the activities associated with a service order, from receiving the customer service requests to the actual service turn-up at the completion of the verification. In case of the LEC, the outside plant department (OSP) and network engineering department carry out these functions. Upon receiving facilities improvement request from another business department, the OSP department begins to work according to following steps: engineering, designing, installing, testing, inspection and turn-up, building in service area, recording.

**(4) Network Engineering Activity**

At the same time, the network engineering department also receives a facilities improvement work order such as transmission equipment and special services (FX circuits, data circuits, etc.) from a business department, which then engineers and plans the implementation.

**(5) Switching and Transmission Managing Activity**

This activity is carried out in the central office department. The typical activities include monitoring, provisioning, maintenance checks, rearrangement, verifying operation of transmission systems, performing fiber optic and multiplexer, transmission tests, and installing and trouble shoot special circuits.

The operation of the switch is done through the Master Control Center (MCC) panel and/or a terminal. Remote operations are also done through these input/output channels. Figure 3-4 shows a sequence of activities to operate and manage an actual LEC's network.

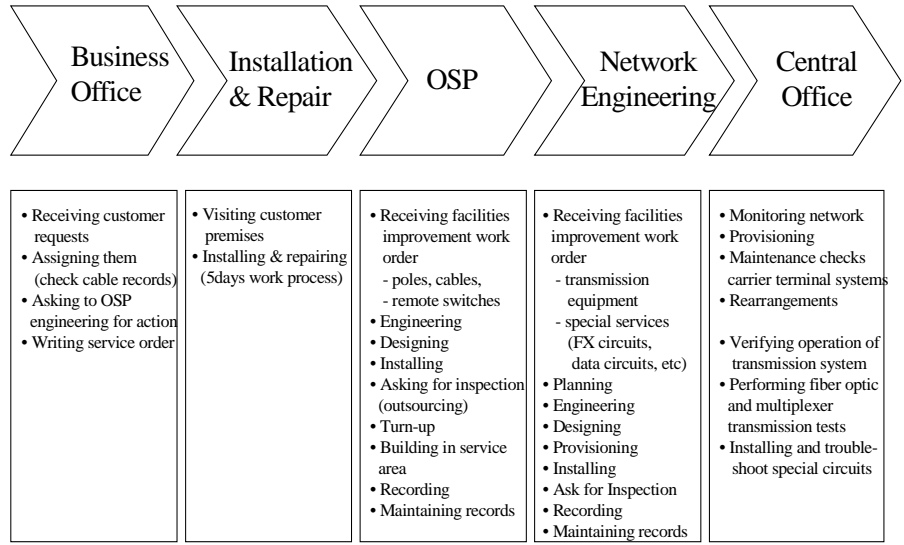

**Figure 3-4 Activity Map of Network Operating**

### 3.2.4.2 NETWORK MANAGEMENT

Figures 3-5 shows the architecture and each components of network management system of the ILEC network as of 2000. As seen in Figure 3-5, the LEC's network management system is not integrated. For example, SONET manger can only control Sonet, etc.. Also the LEC's network management systems cannot interoperate because they consist of different products from different vendors, which complicates network management.

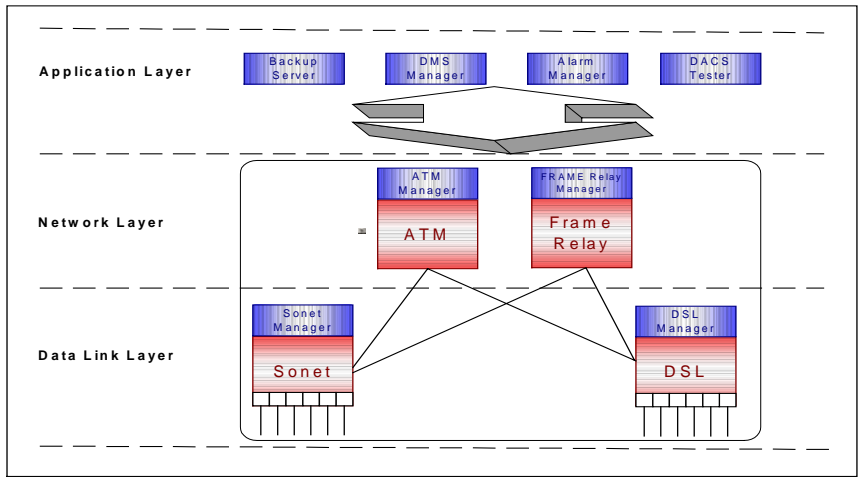

**Figure 3-5 Network Management System of the LEC**

### 3.3  PittNet (University of Pittsburgh IP Network)

#### 3.3.1 Network Description

At present, University of Pittsburgh is converting to a fiber-based Gigabit Ethernet backbone network known as PittNet.  It provides service to 32,000 undergraduate, graduate students, and over 9,000 faculty & staff.  It is comprised of more than 22,000 Ethernet and serial ports and 800 modems.  It is constructed using routers, switches, and other networking equipment from Cisco Systems. A limited number of switches and hubs from other vendors are also supported.  Figure 3-8 shows the architecture of PittNet.

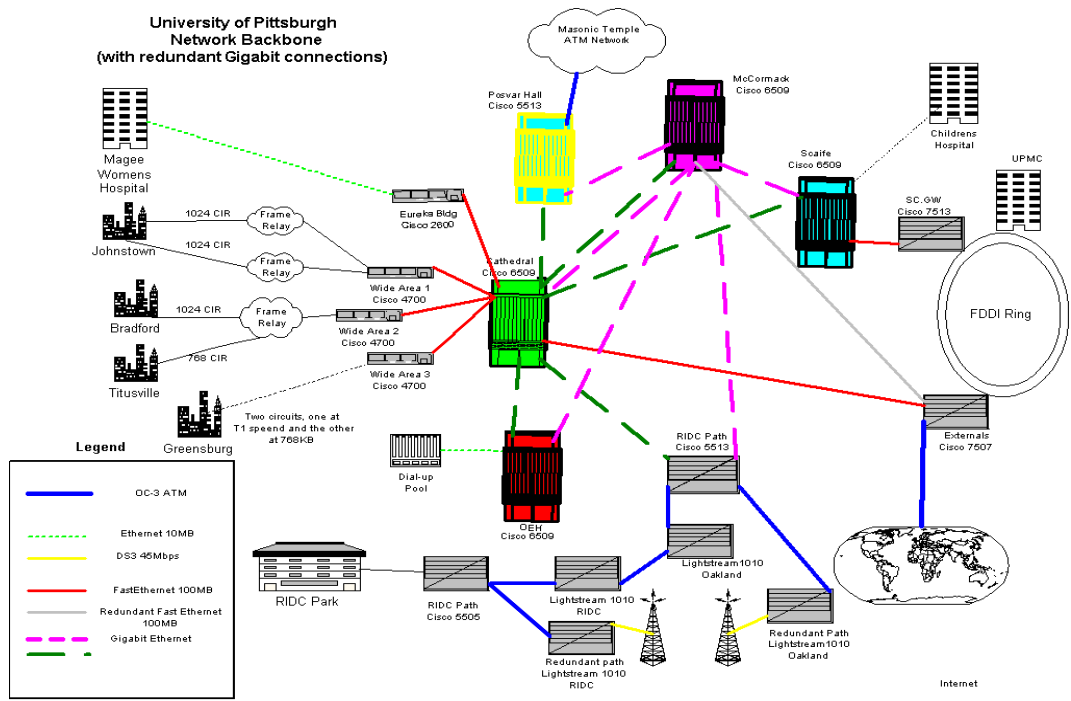

Figure 3-8 Network Architecture of PittNet

The PittNet backbone has the ability to transmit data at a rate of 4 Gbps full duplex.  Most ports connect at 10 Mbps and are an even mix of shared and switched Ethernet.  As of 2000, there were approximately 200 ports connected at 100 Mbps.  All links between routers, core switches, building entrance routers, and building entrance switches are fiber.  If one core switch fails, there is an automatic fail-over to the redundant path.  If a router fails on the Pittsburgh campus, connectivity is lost to the subnets that are fed from that router.  The only redundant routers are in the path to RIDC and to the subnets in the RIDC machine room.  PittNet currently has connections to several other facilities and networks.

The Regional campuses connect to PittNet with frame relay or T1 links.  The Bradford campus is connected by a frame relay link with a 1024 Kbps committed information rate (CIR).  The Greensburg campus is connected by a T1 link that is half voice and half data. An order has been placed for an additional T1 link that will provide one dedicated T1 link for voice and one for data. This configuration is needed at the Greensburg campus to support the ports in the residence halls. The Johnstown campus is connected using a frame relay link with a 1024 Kbps CIR. The Titusville campus is connected using a frame relay link with a 768 Kbps CIR. These low speed connections to the Regional campuses require many bandwidth intensive services to be replicated on the individual campuses.

### 3.3.2 Computing Services and Systems Development (CSSD)

Computing Services and Systems Development (CSSD) provides the network infrastructure and telecommunications backbone for the University. CSSD operates the Technology Help Desk, which is available 24 hours a day, 7 days per week. It provides troubleshooting, problem resolution, and answers to the questions on a variety of information technology issues. Students have access to computing resources and services via six university computing labs.

### 3.3.3 Network Operation and Management (OAM)

### 3.3.3.1 Network Operation

Network operation and customer services include the development of a centralized trouble desk staffed by first-level support staff, who will resolve simple technical questions and route medium or complex problems to appropriate personnel for fast resolution..

### 3.3.3.2 Network Management

Hewlett-Packard's OpenView product has been selected as the network management system for PittNet. Veritas' Nerve Center product has been selected for event correlation. And also Landmark's Performance Works software has been selected and will be deployed across computing platforms.

PittNet's network management is distributed system management. The implementation of network management protocol is SNMP. It is a protocol to manage the same components of the network and to transfer management information over the network. It has also very simple architecture. Adoption of a single protocol network simplifies router configuration and enhance the reliability of the network. The single protocol to be routed is TCP/IP, an international standard for network transport.

## 4. COST ANALYSIS

### 4.1 General Assumptions

We implicitly assume that circuit-switched network (PSTN) has different cost characteristics than packet-switched network (IP network). In particular, we assume the following:

- All the switches and routers in the service areas are equipped with QoS-support functionality.
- Both networks maintain enough redundancy to guarantee reliability of communication (e.g., below 1% blocking rate).
- Geographical distance doesn't have much influence on the network operating costs.
- Software upgrade for network operation and management occurs periodically with a certain interval and its cost occurs proportionally.
- Both networks are managed by appropriately skilled staff to deliver support services efficiently -i.e. not using under- or over-skilled staff.
- Both networks are carried efficient processes for the operation of the support and management function.
- The management tools are appropriate to the operational tasks.

### 4.2 ILEC's Network (PSTN)

Figure 4-1 shows a cost structure of the ILEC in 1999. The cost structure in 1998 and 1997 is similar to that in 1999. As seen, switching and transmission expenses are not higher than non-specific plant expense and customer service is high. Furthermore non-related network expenses are more important than network-related expenses.

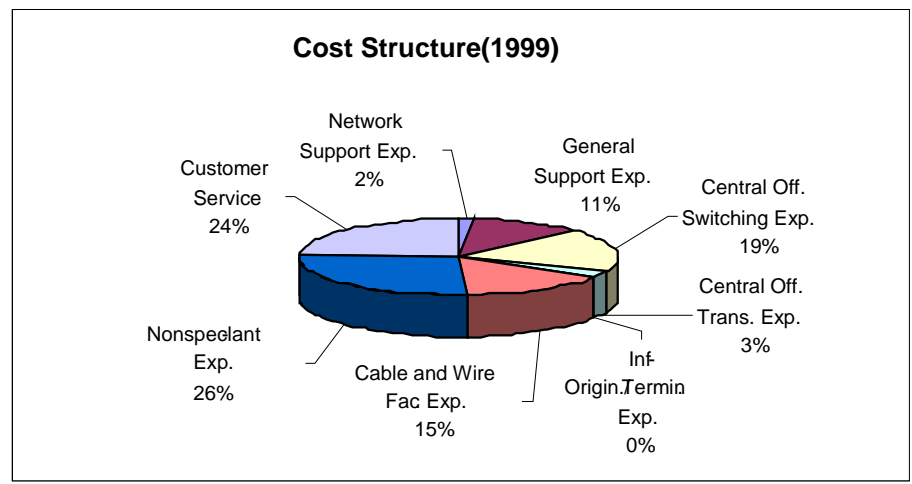

**Figure 4-1 Cost Structure of the ILEC (1999)**

Figure 4-2 shows the traditional telephony industry average cost structure. The result is similar to the ILEC we use as our case. We therefore conclude that the cost structure is stable in circuit-switched networks.

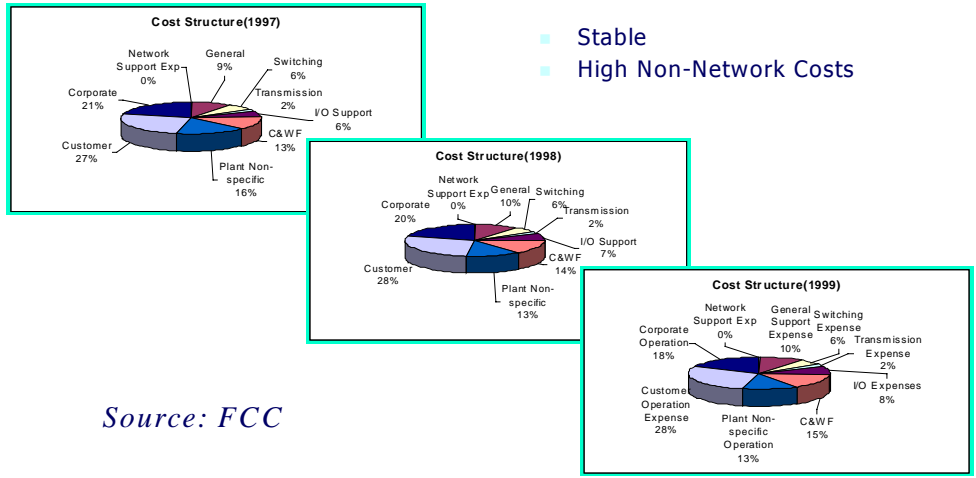

*Source: FCC*

**Figure 4-2 Cost Structure of traditional telephony industry**

Figure 4-3 shows the simple regression result between total operating costs and cost drivers (subscribers, switch lines, and sales) in the traditional telephony industry. There is a strong linearity, which implies that a simple regression model can be useful to estimate operating costs.

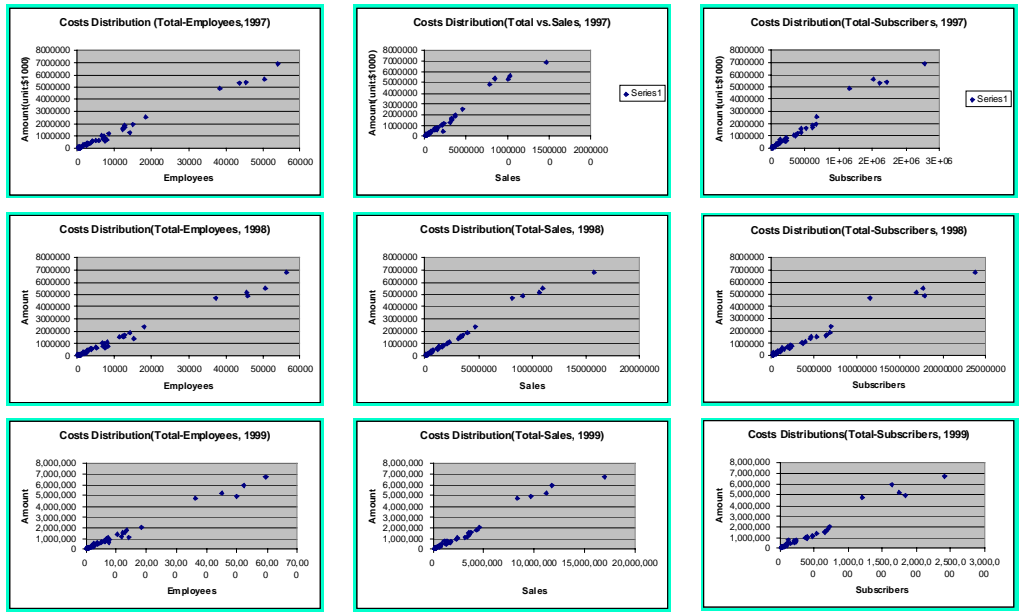

**Figure 4-3 the relationship between cost items and cost drivers**

Table 4-1 shows the average cost per subscriber, per employer, and per sales during three years. We measure the unit average cost by consisting of different groups. In total, 51 companies reported their financial results to the FCC. The "removed outlier" group (46) consists of all telcos *except* for GTE and five RBOCs. Even though they are only six telcos, their influence on the costs are substantial (nearly 50%). We further separated this "removed outlier" group (6) into two groups: the large-size group (24) and small-size group (22) by considering subscriber, employees and sales to investigate whether size influences operating costs. We interpret the results (see table-1) as follows:

- First, the average operating cost between total group (51) and the "removed outlier" group (46) is not very different. The main reason is that the relationship between operating costs and the cost drivers (subscriber, employees, and sales) are purely linear (as seen in the above).
- Second, there is a large difference between large-size group (24) and small-size group (22) in the average operating costs. We called it "size effect'. This size effect occurs because the initial fixed operating costs are large compared to additional operating costs, even though the operating costs are linear with cost drivers.

| Year | | 1997 | 1998 | 1999 | Average |
|---|---|---|---|---|---|
| Total Group (51) | Subscriber | $331 | $308 | $296 | $219 |
|  | Employees | $124,011 | $120,350 | $118,57 | $120,939 |
|  | Sales | $0.52 | $0.48 | $0.46 | $0.49 |
| Removed Outlier (46) | Subscriber | $323 | $307 | $279 | $211 |
|  | Employees | $127,588 | $127,153 | $124,578 | $126,440 |
|  | Sales | $0.49 | $0.48 | $0.43 | $0.47 |
| Large-size Group (24) | Subscriber | $328 | $304 | $292 | $217 |
|  | Employees | $122,703 | $118,657 | $116,127 | $119,163 |
|  | Sales | $0.52 | $0.48 | $0.45 | $0.48 |
| Small-size Group (22) | Subscriber | $361 | $353 | $339 | $245 |
|  | Employees | $140,903 | $142,380 | $149,592 | $144,292 |
|  | Sales | $0.52 | $0.50 | $0.49 | $0.50 |

**Table 4-1 The Average Cost of Traditional Telephony Industry**

Table 4-2 shows the result of estimated operating costs using the regression model, which is estimated using traditional telephony industry data. The cost data of the ILEC under study was put into the estimated regression formula. An ILEC is one of the small-size groups in the above and actual operating costs of the ILEC, as of 1999, was 25 million dollars. As a result, we conclude that the ILEC's operating costs are explained well by subscribers.

| Cost Drivers | Total Group | Removed Outlier | Large-size Group | Small-size Group |
|---|---|---|---|---|
| Subscribers | $21,979,800 | $20,739,480 | $21,711,470 | $25,212,850 |
| Employees | $30,087,950 | $31,642,920 | $29,496,240 | $37,996,420 |
| Sales | $28,232,490 | $26,900,380 | $28,052,430 | $30,246,820 |

**Table 4-2 The Estimated Operating Costs (1999)**

Based on the LEC's cost data, we observe and conclude the following:

- The total operation costs were approximately $25 million per year (three-year average costs from 1997 to 1999).
- Operating costs per subscriber per year is about $337. This amounts to $28 per subscriber per month, and is not differentiated by service type.
- Operating costs per employee per year is about $99,383.

- Equipment oprating costs are about 6% of total equipment costs. Of them, 3.2% is labor costs (1%: transmission and 2.2%: switching) and 2.8% is training, vehicles, software, spare cards, etc.
- Customer service costs are the main components of operating costs. They comprise approximately 24% of total operating costs (except for depreciation costs).
- Network related operation (CO switching and transmission) costs are about 22%.

## 4.3 PittNet (University of Pittsburgh IP Network)

Figure 4-4 shows a cost structure of PittNet in 1999. The cost structure of PittNet is quite different from that of the LEC's network. One significant reason is that the school follows accounting rules of non-profit organizations. Another reason is that, unlike an ILEC, the University is not commercial. So, customer service (student service) portion is very small (only 6%), but the ILEC's is about 24% (the highest portion of operating costs).

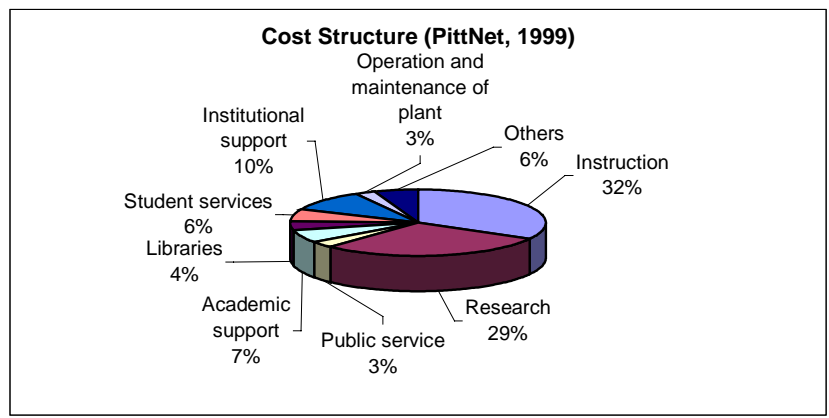

**Figure 4-4 Cost Structure of PittNet (1999)**
*Source: Office of Budget and Financial Reporting, University of Pittsburgh, 2000*

Based on PittNet's cost data, we make the following conclusions and observations:

- The average total costs for the operation and maintenance of plant were approximately $21,620,000 per year (three-year average costs from 1997 to 1999). This ranged between $19,845,000 and $23,292,000 per year, depending on the scale and complexity of the network, and type of network. This is equivalent to approximately $720 per node of total operating costs
- Equipment operating costs for PittNet are lower than those of ILEC. The reasons appear to be that PittNet tends to be a relatively simple LAN-based network.
- Staff costs are the main components of operating costs. They comprise approximately 39% of total operating costs (except for depreciation costs). Further, some of the staff costs were hidden through the use of teaching staff to support the networks on a part-time basis. For PittNet, this amounts to $8.43 million per year and $281 per node.

- Operation and maintenance costs of plant are about 3%. This is equivalent to $648,600 per year, or $21.60 per node.

**4.4 Comparison of the Data**

The three main operating costs in networks are hardware (equipment), software and people (network operation & management). So, next section deals with these three cost drivers for the comparison of both networks, an ILEC PSTN and PittNet.

**4.4.1 Hardware**

The ILEC's switch (DMS SuperNode System) is manufactured by a single vendor who guarantees interoperability among components. Further, although the ILEC has remote switching modules (RSCs -- remote switches and RLCMs -- remote line concentrating modules), the network has a substantially centralized architecture. While routers in PittNet have distributed and decentralized network environment, which is characterized primarily by its topological features and client-server organization. Both the switch-based network (ILEC) and the router-based network (PittNet) are modular in design, but the switch is limited because its architecture is integrated, which means that each module is interdependent. This means that whole in the system must be changed, even if only some modules need to be adapted to a new application or use. This descriptive analysis coincides with the following quantitative results concerning the equipment's maintenance costs, which is related to the network hardware (physical) architecture.

(1) In the case of the ILEC, the average annual maintenance costs of equipment are 6% of total equipment acquisition costs. The ILEC has two switches with an acquisition cost of approximately one million dollars each. So, the maintenance costs for switches in an ILEC is roughly $120,000 per year. (6% × $2,000,000)

(2) In case of PittNet, Table 4.3 shows the maintenance costs of routers, which are based on data provided by the network administrator of PittNet. Catalyst 6500 is used in big hub sites and large building entrances. The 3508 router is used in small building entrances and the 3548 routers are switches that will be in closets, whether the building is large or small. The 2600 router is the smallest router in PittNet.

| Types  | Catalyst 6500 | 3548  | 3508  | 2600  |
|--------|---------------|-------|-------|-------|
| Amount | $6,500        | $475  | $895  | $392  |

**Table 4.3 Maintenance costs of Routers**

As of 1999, Table 4.4 shows the number of router in PittNet. To be conservative, we assume that the maintenance costs of equipment not listed will be about the same as the next higher new-generation equipment (until better data are available). For example, we assume the maintenance costs of Catalyst 5505 will be $6,500. From the above data , the total maintenance costs of the routers used in PittNet is about $78,392, or $2.61 per node. This compares to the equivalent cost for the ILEC of $120,000, or $1.60.

| Types | 7513 | 7507 | 6509 | 5505 | 4700 | 2600 |
|---|---|---|---|---|---|---|
| No. of Routers | 1 | 1 | 4 | 1 | 3 | 1 |

**Table 4.4 The number of routers in PittNet**

### 4.4.2 Software

As the same with hardware, the software in an ILEC is complex, proprietary and integrated architecture. The current average annual software upgrades cost is one million dollars ($1,000,000) in each of the ILEC's switches.

In case of PittNet has an open and layered architecture that is the same as the legacy IP network architecture. We have not yet obtained the exact data because the software costs in PittNet are mixed with general operating software for network operation & management and specific software for students. Further, PittNet will need more software costs for QoS functions in the future applications like VoIP.

### 4.4.3 People

The ILEC has 25 employees related to switching as of 1999. From regulatory data, we compute the average salary to be about $58,814 per person per year. Thus, we estimate that total network management personnel costs is $4,411,050, or $58.80 per station. In PittNet, the average salary of technical employees is about $58,284. There are 6 people working directly on networking issues, resulting in personnel costs of approximately $350,000, or $11.60 per node[1].

### 4.5 Summary of Results

Table 4.5 summarizes the results presented in Section 4.4.  If per-node costs are constant with scale (a *big* assumption that must be validated), the software costs for VoIP are about the same as those for the circuit switched network, and the requisite gateways do not contribute significantly to operating costs, then per-station VoIP costs are $27.54 vs. $75.73 for the ILEC's network.  This is a reduction of approximately 64%, which is quite substantial. Because of the stated caveats, the authors believe that this is a best-case outcome, and that the actual VoIP operating costs will be higher.

|  | ILEC | PittNet |
|---|---|---|
| Hardware | $1.60 | $2.61 |
| Software | $13.33 | ?? |
| People | $58.80 | $11.60 |
| Total | $73.33 | >$14.21 |

**Table 4.3 - Summary of Per-Station Operating Costs for Switching**

There are other mitigating factors that would have to be considered as well, however.  In a transition to VoIP, the customer service costs are likely to be much higher because:

- It is not at all clear that the services and features that VoIP will provide will be identical to those currently provided by the PSTN.  Thus, additional resources will be needed to describe, educate, and assist customers with the new procedures.

---

[1] These figures are estimates that must still be validated.

- Problems are likely to be encountered in the transition that will cause confusion for customers.
- It is likely that some customers will opt for IP telephones and eschew the gateway, while others will continue to use their same customer premise equipment through the gateway. This means that customer service has to be capable of providing support for both classes of devices.

## 5. Implications and Concluding Remarks

Our results are based on a pure substitution scenario (where VoIP simply replaces circuit switching, no more, no less). Thus, the customer service and outside plant factors are largely unaffected (in the steady state) by the difference of switching technology. Further, the substantial majority of a telco's operating cost lies in customer service and outside plant maintenance: of an overall operating cost of approximately $25M, the switching-related costs are roughly 24% ($6M) for the LEC. So even if the cost differences for substitute services were large, the overall impact on the telco's financial performance might not be significant.

In conclusion, we conducted a study focused on the operation of circuit-switched and IP networks especially operational efficiency and operating costs. Our study was implemented by investigating actual networks, the LEC and PittNet. Although our results cannot be generalized to the industry as a whole, it is meaningful as a first step to investigating the actual operating costs of circuit-switched and IP networks. We will continue this research by focusing on the migration issues faced by the LEC. In particular, we will focus on the systems needed to provide quality of service.

DRAFT                                                                                                   DRAFT